\documentclass[preprint,showpacs,preprintnumbers,amsmath,
amssymb,superscriptaddress,floatfix]{revtex4}
\usepackage{graphicx}
\usepackage{epstopdf}
\usepackage{amsmath}
\usepackage{bm}
\usepackage{mathrsfs}
\usepackage{color}
\usepackage{slashed}
\usepackage{dcolumn}
\usepackage{float}

\newcommand{\be}{\begin{equation}}
\newcommand{\ee}{\end{equation}}
\newcommand{\ba}{\begin{eqnarray}}
\newcommand{\ea}{\end{eqnarray}}
\newcommand{\nn}{\nonumber}

\newcommand{\la}{\langle}
\newcommand{\ra}{\rangle}

\newcommand{\RNum}[1]{\uppercase\expandafter{\romannumeral #1\relax}}

\begin{document}

\title{The $K\bar{K}^*$ interaction in the unitary coupled-channel
approximation}

\author{Da-Ming Wan}
\affiliation{College of Applied Sciences, Beijing University of
Technology, Beijing 100124, China}

\author{Si-Yu Zhao}
\affiliation{College of Applied Sciences, Beijing University of
Technology, Beijing 100124, China}

\author{Bao-Xi Sun}
\email{sunbx@bjut.edu.cn} \affiliation{College of Applied Sciences,
Beijing University of Technology, Beijing 100124, China}


\begin{abstract}
The $K\bar{K}^*$ interaction Lagrangian is constructed when the $SU(3)$
hidden gauge symmetry is taken into account, and then the
$K\bar{K}^*$ potential is obtained. In the low energy region, the
$K\bar{K}^*$ potential mainly comes from the contribution of the
$t-$channel interaction by exchanging $\rho$,$\omega$ and $\varphi$
mesons, respectively.
The $K\bar{K}^*$ amplitude is investigated by solving the
Bethe-Salpeter equation in the unitary coupled-channel
approximation, where the loop function of the vector and
pseudoscalar mesons are evaluated in the dimensional regularization
scheme, and the contribution of the longitudinal part of the
intermediate vector meson propagator is included in the calculation.
Finally, it is found that a resonance state of $K\bar{K}^*$ is
generated in the isospin $I=0$ sector, which might correspond to the
$f_1(1420)$ particle in the review of the particle data group(PDG).
Moreover, in the isospin $I=1$ sector, a pole of the $K\bar{K}^*$
amplitude is detected at $1425-i316$MeV on the complex plane of the
total energy in the center of mass system, which is higher than the
$K\bar{K}^*$ threshold. Thus this pole might be a resonance state of
$K\bar{K}^*$ although no counterpart has been found in the PDG
review.
\end{abstract}


\maketitle

\section{Introduction}
\label{sect:Introduction}

Quantum Chromodynamics(QCD) is considered to be the fundamental
theory to describe strong interactions. In the high energy region,
the results obtained with QCD are in good agreement with the
experimental data because of the asymptotic freedom. However, the strong
coupling constant increases rapidly when the energy becomes lower,
and the method of the perturbation expansion will not
be applied, so other methods are developed to deal with the
non-perturbative QCD effect, such as Lattice QCD theory, QCD sum
rules, the constituent quark model, and the chiral unitary method.

In the chiral unitary method, the Bethe-Salpeter equation is solved
while the unitarity of the amplitude is reserved. Thus the resonance
state of the interacting hadrons could be generated dynamically. In
the low energy region, it has achieved great success in the study of
the interaction of hadrons\cite{Kaiser95,Oller97,Ramos97}.

The vector meson component can be introduced in the interaction
Lagrangian when the hidden gauge symmetry is taken into account\cite{16,17,18,19}.
Along this clue, the pseudoscalar meson and vector meson
interaction\cite{20}, the vector meson and vector meson
interaction\cite{21,22},the vector meson and baryon octet
interaction\cite{23,24}, and the vector meson and baryon decuplet
interaction in the SU(3) flavor space are studied in the unitary
coupled-channel approximation\cite{25,26}.

In the present work, the $K\bar{K}^*$ interaction will be
evaluated in the $SU(3)$ flavor space when the hidden gauge
symmetry is considered, and then we will examine whether the resonance
state with a $s\bar{s}$ quark pairs can be generated dynamically by
solving the Bethe-Salpeter equation in the unitary coupled-channel
approximation. The article is organized as follows:
In Section~\ref{sect:KKpotential}, the $K\bar{K}^*$ potential is
constructed in the SU(3) flavor space when the hidden gauge symmetry
is taken into account. In Section~\ref{sect:BS}, the Bethe-Salpeter
equation is discussed in the case of the $K\bar{K}^*$ interaction.
Especially, the unitarity of the amplitude is emphasized when the
longitudinal part of the intermediate vector meson propagator is
included in the calculation. The results are given in
Section~\ref{sect:results}, and it is summarized in
Section~\ref{sect:summary}.

\section{$K\bar{K}^*$ potential}
\label{sect:KKpotential}

According to the hidden gauge symmetry, the interaction Lagrangian
of the vector meson and the pseudoscalar meson can be written as
\be {\cal L}_{VPP}=-i g \la V_\mu [P, \partial^\mu P ] \ra,
\label{L1} \ee
where $g=\frac{M_V}{2f_\pi}$, $f_\pi=93$MeV is the pion decay
constant, $M_V$ is the mass of vector meson, and
\be
P=\left(\begin{array}{ccc}
 \frac{1}{\sqrt{2}}\pi^0+\frac{1}{\sqrt{6}}\eta & \pi^+ & K^+  \\
 \pi^- & -\frac{1}{\sqrt{2}}\pi^0+\frac{1}{\sqrt{6}}\eta &      K^0 \\
K^- & \bar{K}^0 &  -\frac{2}{\sqrt{6}}\eta
                                               \end{array}
                                          \right),
\label{PJZ}
\ee
and
\be
V_\mu=\left(\begin{array}{ccc}
 \frac{\omega}{\sqrt{2}}+\frac{\rho^0}{\sqrt{2}} & \rho^+ & K^{*+}  \\
 \rho^-                                          & \frac{\omega}{\sqrt{2}}-\frac{\rho^0}{\sqrt{2}} & K^{*0} \\
  K^{*-}                                         & \bar{K}^{*0}                                         &  \varphi
                                               \end{array}
                                          \right)_\mu
\label{VJZ}
\ee
 are the pseudoscalar meson
matrix and the vector meson matrix in the $SU(3)$ flavor space, respectively\cite{20}.

The vector meson Lagrangian takes the form of
\ba
\label{eq:VmunuVmunu}
 \mathcal{L}_V=-\frac{1}{4}\la V_{\mu\nu}V^{\mu\nu} \ra,
\ea
where
\ba
V_{\mu\nu}=\partial_{\mu}V_{\nu}-\partial_{\nu}V_{\mu}-ig[V_{\mu},V_{\nu}]
. \label{VVV} \ea

According to Eqs.~(\ref{eq:VmunuVmunu}) and (\ref{VVV}), the
interaction Lagrangian of three vector mesons can be written as
\ba \mathcal{L}_{VVV}=ig \la
(\partial_{\mu}V_{\nu}-\partial_{\nu}V_{\mu})V^{\mu}V^{\nu} \rangle
. \label{L2} \ea
Thus the $\bar{K}^*\bar{K}^*\rho$, $\bar{K}^*\bar{K}^*\omega$, and
$\bar{K}^*\bar{K}^*\varphi$ vertices can be obtained easily.

\begin{figure}[H]
\centering
\includegraphics[scale=0.5]{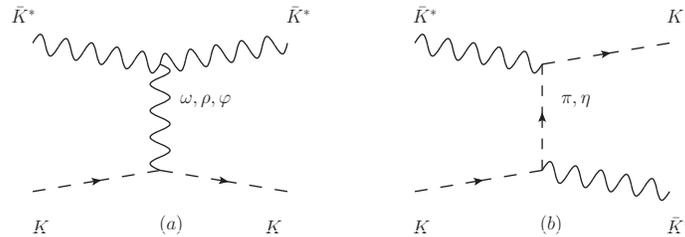}
\caption{The interaction of $\bar{K}^*K\rightarrow \bar{K}^*K$, (a)
Vector meson exchange, (b) $\pi$ and $\eta$ exchange.} \label{t1}
\end{figure}

The $K\bar{K}^*$ potential is composed of $t-$ channel and $u-$
channel interactions, as shown in Fig.~\ref{t1}. For the $t-$
channel interaction, there are three intermediate mesons exchanged,
i.e., $\omega$, $\rho$ and $\varphi$, respectively.
In this work, we study the scattering of $K\bar{K}^*$ in the low
energy region, so the momentum of the intermediate meson is ignored.
From the Feynman diagram in Fig.\ref{t1}(a), we can see that the
$t-$channel interaction of $K\bar{K}^*$ can be divided into two
parts, and the one is the vertex of $\bar{K}^*\bar{K}^*$ and a
vector meson, and the other is the vertex of $\bar{K}K$ and a vector
meson. Therefore, we can calculate these two kinds of vertices and
then combine them together to obtain the $t-$ channel potential of
$K\bar{K}^*$, which can be written as
\ba \label{eq:201808161618}
 V_{ij}&=&C_{ij}\frac{1}{f^2_{\pi}}(p_1+p_2)\cdot(k_1+k_2)
\varepsilon\cdot\varepsilon^* \nonumber \\
&=&\tilde{V}_{ij}g^{\mu \nu} \varepsilon_\mu \varepsilon^*_\nu,
 \ea
with \be \label{eq:8171602}
\tilde{V}_{ij}=C_{ij}\frac{1}{f^2_{\pi}}(p_1+p_2)\cdot(k_1+k_2). \ee
In Eq.~(\ref{eq:201808161618}), the label $i$ denotes the initial
state and the label $j$ stands for the final state, $p_1(k_1)$ and
$p_2(k_2)$ represent the initial and final momenta of the
$\bar{K}^*(K)$ meson, and $\varepsilon$ and $\varepsilon^*$ are
polarization vectors of the initial and final $\bar{K}^*$ mesons,
respectively.
Since the values of $\rho$, $\omega$ and $\varphi$ meson masses are
similar to each other, we assume that $M_V\approx m_{\rho}\approx
m_{\omega}\approx m_{\varphi}$, then the coefficients $C_{ij}$ of
the $K\bar{K}^*$ potential take the simple values as listed in
Table~\ref{bKV}.
\begin{table}[H]
\centering
\begin{tabular}{|c|c|c|c|c|}
\hline
      $C_{ij}$    &$\bar{K}^{*0}K^0$ &$K^{*-}K^+$ &$K^{*+}K^-$ &$K^{*0}\bar{K}^0$\\
\hline
$\bar{K}^{*0}K^0$ &$0.5$ & $0.25$& 0& 0\\
\hline
$K^{*-}K^+$ &$0.25$ & $0.5$ & 0&0\\
\hline
$K^{*+}K^-$ & 0 & 0& $0.5$ &$0.25$\\
\hline
$K^{*0}\bar{K}^0$ & 0& 0 &$0.25$ &$0.5$\\
\hline
\end{tabular}
\caption{The coefficients $C_{ij}$ in the $K\bar{K}^*$ interaction,
$C_{ji}=C_{ij}$.} \label{bKV}
\end{table}

If Mandelstam variables $s=(p_1+k_1)^2$, $t=(k_2-k_1)^2$ and
$u=(p_2-k_1)^2$ are adopted, the kernel in Eq.~(\ref{eq:8171602})
can be written as \ba \label{eq:tildeV1222222}
\tilde{V}_{ij}&=&C_{ij}\frac{1}{f^2_{\pi}} (s-u) \nn \\
&=&C_{ij}\frac{1}{f^2_{\pi}} \left( 2s+t-2(M_K^2+M_{K^*}^2) \right).
\ea
Around the $K\bar{K}^*$ threshold, Mandelstam variable
$t=(k_2-k_1)^2$ is assumed to be zero, so
Eq.~(\ref{eq:tildeV1222222}) can be simplified as
\ba \label{eq:tildeV1222333} \tilde{V}_{ij}
&=&C_{ij}\frac{2}{f^2_{\pi}} \left( s-M_K^2-M_{K^*}^2 \right). \ea

According to the interaction Lagrangian of the vector meson and the
pseudoscalar meson in Eq.~(\ref{L1}), the $u-$ channel potential of
the $K\bar{K}^*$ interaction via a pion or an $\eta$ meson exchange
in Fig.~\ref{t1}b can be obtained as
\ba
V_{\alpha ij}&=&D_{\alpha ij}g^2(q-k_1)\cdot \varepsilon^*
\frac{1}{q^2-m^2_{\alpha}}(q-k_2)\cdot \varepsilon \nonumber\\
&=&4 D_{\alpha ij}g^2 q\cdot
\varepsilon^*\frac{1}{q^2-m^2_{\alpha}}q\cdot \varepsilon,
\ea
where $\alpha$ represents $\pi$ or $\eta$, $D_{\alpha ij}$ is the
interaction coefficient, $q=p_2-k_1=p_1-k_2$ is the momentum of the
intermediate meson, $p_1\cdot \varepsilon=0$ and $p_2
\cdot\varepsilon^*=0$.
The zero component of the polarization vector of the $\bar{K}^*$
meson is in inverse proportion to the $\bar{K}^*$ meson mass, thus
it can be neglected, so we have
\be \label{eq:qdote} q \cdot \varepsilon^* \sim |\vec{q}|, \ee and
\be \label{eq:qdotepr} q \cdot \varepsilon \sim |\vec{q}|. \ee
If the three-momentum of the initial and final mesons is neglected,
just as we have done in this work and other works of ours, the $u-$
channel potential of $K\bar{K}^*$ is trivial. Therefore, only the
the $t-$ channel potential of the $K\bar{K}^*$ interaction is taken
into account when the Bethe-Salpeter equation is solved.

In addition to the $K\bar{K}^*$ channel, there are other channels
with strangeness zero, i.e., $\pi \rho$, $\pi \omega$, $\pi
\varphi$, $\eta \rho$, $\eta \omega$, $\eta \varphi$. However, the
$\pi \rho$ threshold is far lower than the energy region where we
are interested, and the elastic scattering amplitude of the other
five channels are all zero. Thus the influence of these six channels
are neglected when the $K\bar{K}^*$ interaction is studied in this
work.

\section{The Bethe-Salpeter equation}
\label{sect:BS}

The Bethe-Salpeter equation can be written as
\ba \label{eq:Behtesalpeter}
T&=&V+VGV+VGVGV+... \nonumber \\
 &=&V+VGT,
\ea
where $T$ is the scattering amplitude, $V$ is the interaction
kernel, and $G$ is the loop function, which is a diagonal matrix.
In the interaction of the pseudoscalar meson and the vector meson,
the loop diagram in the Bethe-Salpeter equation can be written as
\ba \label{Gll}
G_l(s)&=&i\int \frac{d^4 q}{(2\pi)^4} \frac{-g_{\mu \nu}+\frac{q_\mu q_\nu}{M_a^2}}{q^2-M_a^2+i\varepsilon}\frac{1}{(P-q)^2-M_b^2+i\varepsilon}\nonumber\\
&=&-g_{\mu\nu}G_{ab}(s)-\frac{g^{\mu\nu}}{M_a^2}H^{00}_{ab}(s)-\frac{P^{\mu}P^{\nu}}{M^2_a}H^{11}_{ab}(s),
\ea
where $M_a$ and $M_b$ are masses of the vector meson and the
pseudoscalar meson, respectively, and $P=p_1+k_1=p_2+k_2$ is the
total momentum of the system. The term relevant to the transverse
part of the propagator of the intermediate vector meson is
 \ba G_{ab}(s)=i\int \frac{d^4
q}{(2\pi)^4}\frac{1}{q^2-M_a^2+i\varepsilon}\frac{1}{(P-q)^2-M_b^2+i\varepsilon},
\ea and the functions $H^{00}_{ab}(s)$ and $H^{11}_{ab}(s)$
correspond to the contribution of the longitudinal part of the
propagator of the intermediate vector meson,
 \ba \label{eqqq} g^{\mu \nu} H^{00}_{ab}(s)+P^\mu P^\nu
H^{11}_{ab}(s)=\frac{1}{i} \int \frac{d^4 q}{(2\pi)^4} \frac{q^\mu
q^\nu}{(q^2-M_a^2+i\varepsilon)[(P-q)^2-M_b^2+i\varepsilon]}. \ea

In the dimensional regularization scheme, the function $G_{ab}(s)$
takes the form of \ba
 \label{eq:Gpr}
G_{ab}(s)&=&i \int \frac{d^4 q}{(2\pi)^4} \frac{1}{q^2-M_a^2+i\epsilon } \frac{1}{(P-q)^2-M_b^2+i\epsilon} \nn \\
&=& \frac{1}{16 \pi^2} \left\{ a_l(\mu) + \ln \frac{M_a^2}{\mu^2} +
\frac{M_b^2-M_a^2 + s}{2s} \ln \frac{M_b^2}{M_a^2} + \right.
\nonumber \\ & &  \phantom{\frac{2 M_a}{16 \pi^2}} +
\frac{\bar{q}_l}{\sqrt{s}} \left[
\ln(s-(M_a^2-M_b^2)+2\bar{q}_l\sqrt{s})+
\ln(s+(M_a^2-M_b^2)+2\bar{q}_l\sqrt{s}) \right.   \\
& & \left. \phantom{\frac{2 M_a}{16 \pi^2} +
\frac{\bar{q}_l}{\sqrt{s}}} \left. \hspace*{-0.3cm}-
\ln(-s+(M_a^2-M_b^2)+2\bar{q}_l\sqrt{s})-
\ln(-s-(M_a^2-M_b^2)+2\bar{q}_l\sqrt{s}) \right] \right\}, \nonumber
\ea with the square of the total energy of the system $s=P^2$ and
the three-momentum of the intermediate particles in the center of
mass frame
\begin{equation}
\bar{q}_l=\frac{\sqrt{s-(M_a+M_b)^2}\sqrt{s-(M_a-M_b)^2}}{2\sqrt{s}}.
\end{equation}
The forms of $H^{00}_{ab}(s)$ and $H^{11}_{ab}(s)$ can be found in
the appendix part of Ref.~\cite{Zc3900,X3872}.

The loop function $G_l(s)$ can be written as
\be
\label{eq:201808161636}
 G_l(s)=g^{\mu\nu}\tilde{G}_l(s),
\ee
with
\be
\label{eq:20170820}
\tilde{G}_l(s)=-\left(
G_{ab}(s)+\frac{1}{M_a^2}H^{00}_{ab}(s)+\frac{s}{4M_a^2}H^{11}_{ab}(s)\right),
\ee
approximately.

Replacing the potential in Eq.~(\ref{eq:201808161618}) and the loop
function in Eq.~(\ref{eq:201808161636}) into the Bethe-Salpeter
equation, we obtain
\be \tilde{T}g^{\mu \nu}=\tilde{V}g^{\mu
\nu}+\tilde{V}g^{\mu \alpha}~g_{\alpha
\beta}\tilde{G}~\tilde{V}g^{\beta \nu}+..., \label{eq:Bethe} \ee and
thus \ba
\tilde{T}&=&\tilde{V}+\tilde{V}\tilde{G}\tilde{V}+... \nn \\
&=&[1-\tilde{V}\tilde{G}]^{-1} \tilde{V}. \label{eq:Bethe} \ea The
amplitude $\tilde{T}$ is unitary when the Bethe-Salpeter equation is
solved.

In addition, if the effect of the $\bar{K}^*$ decay width is taken
into account in the calculation, the loop function $\tilde{G}_l(s)$
must be replaced by $\tilde{\tilde{G}}_l(s)$ obtained as
\be
\label{eq:1808171229}
\tilde{\tilde{G}}_l(s)=\frac{1}{N}\int_{(M_a-2\Gamma)^2}^{(M_a+2\Gamma)^2}
d \tilde{m}^2~\frac{1}{\pi}~\frac{M_a
\Gamma}{(\tilde{m}^2-M_a^2)^2+M_a^2 \Gamma^2}
\tilde{G}_l(s,\tilde{m}^2,M_b^2),
\ee
with
\be
N=\int_{(M_a-2\Gamma)^2}^{(M_a+2\Gamma)^2}
d \tilde{m}^2~\frac{1}{\pi}~\frac{M_a
\Gamma}{(\tilde{m}^2-M_a^2)^2+M_a^2 \Gamma^2},
\ee
where the $\bar{K}^*$ decay width $\Gamma=50$MeV for the process of
$\bar{K}^* \rightarrow \bar{K}\pi$,
and the equation
\be
\delta(\tilde{m}^2-M_a^2)=\lim_{\Gamma \rightarrow 0^+}
~\frac{1}{\pi}~\frac{M_a \Gamma}{(\tilde{m}^2-M_a^2)^2+M_a^2
\Gamma^2}
\ee is used.

\section{Results}
\label{sect:results}

The $K\bar{K}^*$ pair with isospin $I=0$ takes the form of
\ba \label{eq:1808171551}
 |K\bar{K}^*\rangle =\frac{1}{\sqrt{4}}(|\bar{K}^0K^{*0}\rangle
+|K^-K^{*+}\rangle -|K^0\bar{K}^{*0}\rangle -|K^+K^{*-}\rangle ),
\ea with $K^{-}=-|1/2,-1/2\rangle$ and $K^{*-}=-|1/2,-1/2\rangle$,
 where the C-parity of the $K\bar{K}^*$ pair is assumed to be
positive since $CKC^{-1}=K^{\dag}$ and $CK^*C^{-1}=-K^{*\dag}$.

According to Eqs.~(\ref{eq:tildeV1222333}) and
(\ref{eq:1808171551}), the kernel of the $K\bar{K}^*$ interaction in
the isospin $I=0$ sector can be written as \ba \label{eq:8171650}
\tilde{V}^{I=0}_{K\bar{K}^*\rightarrow K\bar{K}^*}
&=&\frac{3}{2}~\frac{1}{f^2_{\pi}}(s-M^2_K-M^2_{K^*}) . \ea
Actually, since only non-strange vector mesons act as the
intermediate particles in the $t-$ channel $K\bar{K}^*$ interaction,
the kernel in Eq.~(\ref{eq:8171650}) is independent on the C-parity
of the $K\bar{K}^*$ pair according to the coefficients listed in
Table~\ref{bKV}.
Therefore, the Bethe-Salpeter equation can be solved. Assuming the
subtraction constant $a=-2$ and the regularization scale
$\mu=600$MeV, a pole of the squared amplitude $|T|^2$ appears at
$1394-i83$MeV on the complex energy plane of $\sqrt{s}$, which is
higher than the $K\bar{K}^*$ threshold, and lies in the second
Riemann sheet. This pole can be regarded as a $K\bar{K}^*$ resonance
state, and it is more possible to correspond to the $f_1(1420)$
particle in the review of the Particle Data Group\cite{PDG}, whose
mass and decay width are
$m=1426.4\pm0.9$MeV and $\Gamma=54.9\pm2.6$MeV, respectively.
According to the formula to include the $\bar{K}^*$ decay width in
Ref.~\cite{23}, we obtain a pole at the position of $1394-i75$MeV on
the complex energy plane, which is similar to the result calculated
with Eq.~(\ref{eq:1808171229}).

In the isospin $I=1$ sector, the $K\bar{K}^*$ wave function with the
positive C-parity is constructed as
\ba |K\bar{K}^{*}\rangle =\sqrt{\frac{1}{4}}(|\bar{K}^0K^{*0}\rangle
-|K^-K^{*+}\rangle -|K^0\bar{K}^{*0}\rangle +|K^+K^{*-}\rangle ).
\ea Similarly, the $K\bar{K}^*$ kernel with isospin $I=1$ can be
written as
\ba
\label{eq:8211947} \tilde{V}^{I=1}_{K\bar{K}^*\rightarrow
K\bar{K}^*} &=&\frac{1}{2}~\frac{1}{f^2_{\pi}}(s-M^2_K-M^2_{K^*}) .
\ea
With the kernel in Eq.~(\ref{eq:8211947}), a pole of the squared
amplitude is generated at the position of $1425-i316$MeV on the
complex energy plane of $\sqrt{s}$. Since it is above the
$K\bar{K}^*$ threshold, it can be regarded as a $K\bar{K}^*$
resonance state with isospin $I=1$ although its decay width is too
large and no counterpart is found in the PDG review.
If the decay width of the $\bar{K}^*$ is taken into account with the
formula in Ref.~\cite{23}, the pole is detected at $1432-i330$MeV on
the complex energy plane of $\sqrt{s}$, and the decay width of this
resonance state is still large.

\section{Summary}
\label{sect:summary}

The $K\bar{K}^*$ interaction is discussed when the hidden gauge
symmetry is taken into account, and it is found that the $t-$
channel potential via a non-strange vector meson exchange plays an
important role in the $K\bar{K}^*$ interaction.
The Bethe-Salpeter equation is solved in the isospin space, and the
contribution of the longitudinal part of the propagator of the
intermediate vector meson is taken into account.
In the isospin $I=0$ sector, a resonance state is generated at
$1394-i83$MeV on the complex energy plane, which is above the
$K\bar{K}^*$ threshold, and might correspond to the $f_1(1420)$
particle in the PDG review.
In the isospin $I=1$ sector, a resonance state is generated at
$1425-i316$MeV on the complex energy plane. The decay width of this
resonance is too large, and it implies that this resonance is
unstable. No counterpart is found in the PDG data.
Moreover, the effect of the decay width of the intermediate
$\bar{K}^*$ meson is included in the calculation with two kinds of
formulas, and it manifests that the results are similar to each
other.

\begin{acknowledgments}
Bao-Xi Sun would like to thank Han-Qing Zheng for useful
discussions.
\end{acknowledgments}

\end{document}